\documentclass[twocolumn,showpacs,preprintnumbers,amsmath,amssymb]{revtex4}
\usepackage{graphicx}
\usepackage{dcolumn}
\usepackage{bm}

\begin{document}
\title{Nonlinear dynamics of trapped waves on jet currents and rogue waves}
\author{V.I. Shrira$^1$}
\author{A.V. Slunyaev$^{2,3}$}

\affiliation{$^1$Department of Mathematics, EPSAM, Keele University, Keele ST5 5BG, UK\\
$^2$Institute of Applied Physics, 46 Ulyanova Street, N. Novgorod 603950, Russia, \\
$^3$ Nizhny Novgorod State Technical University, 24 Minina Street, N. Novgorod 603950, Russia}

\date{\today}

\begin{abstract}
 Nonlinear dynamics of surface gravity waves trapped by an opposing jet current is studied analytically and  numerically. For wave fields narrowband in frequency but not necessarily with narrow angular distributions the developed  asymptotic weakly nonlinear theory based on the modal approach of \cite{ShriraSlunyaev2013} leads to the one-dimensional modified nonlinear Schr\"{o}dinger equation of self-focusing type for  a single mode. Its  solutions such as envelope solitons and breathers are considered to be prototypes of rogue waves; these solutions, in contrast to waves in the absence of currents, are robust with respect to transverse perturbations, which suggests potentially higher probability of rogue waves.  Robustness of the long-lived analytical solutions  in form of the  modulated trapped waves and solitary wave groups is verified by direct numerical simulations of potential Euler equations.
\end{abstract}

\maketitle
\emph{Introduction.---}
In the last two decades there was a surge in interest in the phenomenon of extreme or  rogue waves in various areas of physics, e.g. \cite{Onoratoetal2013, YeomEggleton2007, Kharifetal2009}. In the most studied area of water waves in the ocean the main  thrust of the studies was onto the search of mechanisms of rogue waves in the absence of currents \cite{Kharifetal2009}. The modulational (or Benjamin-Feir, BF) instability of narrowband  wave fields has been identified as the mechanism  leading to  formation of anomalously high waves and  significant increase of  their probability. This route has been most intensively studied and is most effective for one-dimensional  (1-d) wave propagation; the theoretical and experimental modeling is  the simplest. However, in reality the 1-d patterns are transversally unstable and, hence, short-lived; there is a dramatic difference in the probability of rogue events due to the BF instability for the strictly 1-d and 2-d  wave propagation: the likelihood of rogue events is much higher for the 1-d  propagation, moreover it  completely vanishes for   wave  fields with the   angular  spectra width exceeding a certain threshold \cite{Onoratoetal2009, Kharifetal2009}. 
On the other hand, it is known that the rogue waves are much more frequent on currents; the Agulhas current gained notoriety in this respect  \cite{Kharifetal2009, Mallory1974}.

To explain increased probability of rogue waves on currents 
the prevailing approach exploits separation of scales between the typical wavelength and current, which leads to the WKB or ray description with a special consideration of  caustics, see the literature reviews in \cite{ShriraSlunyaev2013, Kharifetal2009, HjelmervikTrulsen2009}.

With the focus on nonlinear dynamics, various versions of nonlinear Schr\"odinger equations (NLSE) were derived and analyzed under general assumptions of slow current, weak nonlinearity and narrow-banded spectrum, see \cite{HjelmervikTrulsen2009} and references there.
In  these works the BF instability  was found  to be  strengthened for waves on adverse intensifying current.  The  triggering of the BF instability of a narrow band fields due to the intensification of the current was considered in \cite{JanssenHerbers2009}. However there is an essential feature not captured by the existing NLS-type models: waves  propagating upstream can be trapped by the current, there are multiple caustics. Such trapped waves have been observed on the Gulf Stream and were found to have considerably higher steepness than free waves on current \cite{Kudryavtsevetal1995}. A more general and profound difficulty
 is that there is no technique enabling one to describe wave resonant interactions on currents; the waves refract on currents and hence vary in  space, while the resonant interactions and, in particular, the  resonance conditions have to be described in the wave-vector space. A new approach suggested in  \cite{ShriraSlunyaev2013} allows to overcome these obstacles for currents with a symmetry, in particular, for the common parallel  jet currents.  Instead of operating with rays we deal with the modes  propagating on jet currents for which the standard nonlinear theory  applies.

Here, based on the modal approach and weakly nonlinear  asymptotic expansions, we derive  equations governing one-dimensional   wave evolution along the current; the transverse structure of the field is being  provided by the modes. For a one mode we derive 1-d NLSE without the constraint of a too narrow angular spectra. In contrast to unguided waves, here  the NLSE solitary wave type solutions  are robust. The robustness of such wave patterns suggests dramatic increase of the probability of rogue waves. The predictions of the asymptotic model are validated by direct numerical simulations of the Euler  equations.

\emph{Modal representation for waves on jet currents.---}
 We consider wave motions on the free surface of an ideal incompressible fluid of unit density, waves are propagating along the $Ox$ direction on a given vertically uniform unidirectional steady current $U = \{U(y),0,0\}$.
 The motions are governed by the standard Euler and continuity equations  in  the  domain occupied by the fluid  $z \leq \eta$, where $\eta(x,y,t)$ is the water surface elevation; the water depth is assumed infinite for convenience. These equations  are complemented with the standard boundary conditions for gravity waves: dynamic and kinematic boundary conditions on the surface and decay of velocities as $z \rightarrow -\infty$. We focus on the evolution of waves trapped by the current; trapped modes are selected by the boundary conditions stipulating lateral decay of velocities at $y \rightarrow \pm \infty$.

%
In the linear setting the  problem formulated above was thoroughly  examined in \cite{ShriraSlunyaev2013}. Making use of uniformity of the problem with respect to $x$ and $t$, the Fourier transform may be applied with respect to these variables; then the complete linear solution has the form of a superposition of traveling waves propagating collinear to the current with some structure in $(y,z)$ plane, $w = \Re{ \sum_n{\hat{w}_n(y,z) \exp{\left( i \omega_n t - i k x \right)}} }$, where $w(x,y,z,t)$ is the vertical component of fluid velocity, index $n$ numerates the lateral modes. The modes $\hat{w}_n(y,z)$ are specified  by the two-dimensional boundary value problem (BVP)
\begin{align}\label{2DBVP}
\frac{\partial ^2 \hat{w}_n}{\partial z^2} + \frac{\partial ^2
\hat{w}_n}{\partial y^2}  + \left( \frac{\Omega_n^{\prime\prime}}{\Omega_n}
- 2 \frac{\Omega_n^{\prime 2}}{\Omega_n^2} - k^2 \right) \hat{w}_n = 0 \,, \\
\hat{w}_n|_{y \rightarrow \pm \infty}\rightarrow 0 \,,
\qquad
\hat{w}_n|_{z \rightarrow - \infty}\rightarrow 0 \,, \nonumber
\end{align}
with $\Omega_n(y) = \omega_n - k U$. %
Each mode  is characterized by its cyclic frequency, $\omega_n$, and longitudinal wavenumber, $k$; we choose $k>0$ with no loss of generality.
%

The 2-d BVP (\ref{2DBVP}) may be either solved numerically or reduced to a one-dimensional BVP using an asymptotic  separation of variables  \cite
{ShriraSlunyaev2013}. Here we adopt the second route, which assumes the following representation
$\hat{w}_n = B_n Y_n(y) Z_n(z, y)$. Thus, the mode is specified
by two real functions: $Z_n(z,y)$ determines the vertical structure which depends on $y$ slowly,  $Y_n \partial Z_n / \partial y \ll dY_n / dy Z_n$, and is equal to one on the surface;  $Y_n(y)$ determines the mode transverse dependence. Constants $B_n$ may be complex.

Asymptotic 1-d  reductions of BVP  (\ref{2DBVP}) were derived in \cite{ShriraSlunyaev2013} for two regimes: of `weak' currents (compared to the wave phase velocity), and of `broad' currents (compared to the longitudinal wave length). For the dominant wind waves and swell we are primarily interested in,  all oceanic currents are weak in this sense. The corresponding 1-d BVP  is of the Sturm - Liouville type,
\begin{align}  \label{BVP}
\frac{d^2 Y_n}{d y^2}  + 4 k^2 \left[ \frac{\omega_n}{\omega_g} - \left( 1 + \frac{kU}{\omega_g} \right) \right] Y_n =0 \,, \quad Y_n|_{y \rightarrow \pm \infty}\rightarrow 0\,.
\end{align}
Here $\omega_g = \sqrt{k g}$ denotes the frequency of linear gravity waves unaffected by the current. The wave frequency $\omega_n$ is the to-be-defined eigenvalue of the problem.
For single-humped  currents trapped modes require $kU < 0$ to exist; waves must run against  the current.


\emph{Weakly nonlinear model for modulated waves on jet currents.---}
 Now we concentrate on the regime where nonlinear interactions only between the trapped modes are essential. 
 Employing a standard asymptotic procedure based upon  a small parameter 
 characterizing smallness of wave steepness: $\epsilon \sim k \max{|\eta|}$ or, equivalently, smallness of  fluid velocities, $\epsilon \sim k \max{|w|} / \omega_g$, it is straightforward to derive a variety of evolution equations governing  wave fields nonlinear dynamics; the resulting equations for constants $B_n$ now being  slow functions of time and coordinate $x$ are determined by the choice of  initial configurations of the field. In view of our interest in rogue wave formation we focus on narrowband spectra and consider  wave fields with   the longitudinal spectrum confined to  $\epsilon$ vicinity of the carrier wavenumber, $k=k_c$.

 We stress that  jet currents profoundly modify the picture of wave resonances comparing to the freely propagating deep-water gravity waves; crucially,   three-wave resonances become possible, which results in dynamical equations of the three-wave-interaction type. However, it can be shown that 
  in the limit of weak currents, in which we are primarily interested, the  four-wave interactions  dominate. A key feature of the four-wave regimes is that to leading order wave dynamics is potential. Detailed analysis of all the regimes will be reported elsewhere.  Here, on skipping the derivation, we  provide and briefly discuss the new version of the NLSE which we obtain for the wave fields belonging to a single trapped mode with spectra narrow in  longitudinal wavenumbers $k$; it reads:  
%
%
\begin{eqnarray} \label{NLS}
     - i \left( \frac{\partial A}{\partial t} +  V \frac{\partial A}{\partial x} \right) + \frac{\omega_n}{8k_c^2} \frac{\partial^2 A}{\partial x^2} + \kappa_n  \frac{\omega_n k_c^2}{2}  A |A|^2 = 0, \\
    V=\int_{-\infty}^{\infty}{\left( \frac{k_c g^2}{2 \Omega_n^3} + U \right) Y_n^2 dy} \,,
     \quad
     \kappa_n = \frac{  \int_{-\infty}^{\infty}{  Y_n^4 dy}}{  \int_{-\infty}^{\infty}{Y_n^2 dy}}. \nonumber
\end{eqnarray}
Equation (\ref{NLS}) describes  evolution of the surface elevation, $\eta = \Re{ \left[ A(x,t) Y_n(y) \exp{\left( i \omega_n t - i k_c x \right)} \right]}$, $A$  is linked to the vertical velocity component as $B = i \omega_n A$; the subscripts for $A$ and $B$ are omitted for brevity. Sets of similar
coupled equations appear when more than one mode is initially excited. The  detailed derivation of (\ref{NLS}) and its coupled generalizations will be reported elsewhere.

Equation (\ref{NLS}) differs from the classical NLSE in the still water by the account for the Doppler frequency shift  and a reduced 
 nonlinear coefficient  due to factor $\kappa_n$. By virtue of the  Cauchy--Schwarz inequality  $\kappa_n <1$.
The NLSE  (\ref{NLS}) is of focusing type, hence it supports the BF instability and being integrable 
 it admits a wide class of well studied exact solutions of variable degree of complexity \cite{Osborne2010}. The basic solutions are uniform waves, localized solitary wave groups (envelope solitons) and breathers. Being a one-dimensional evolution equation, (\ref{NLS}) provides a dramatic simplification of  description of complicated dynamics of 2-d nonlinear trapped wave patterns.

\begin{figure}
\centerline{\includegraphics[width=9.5cm]{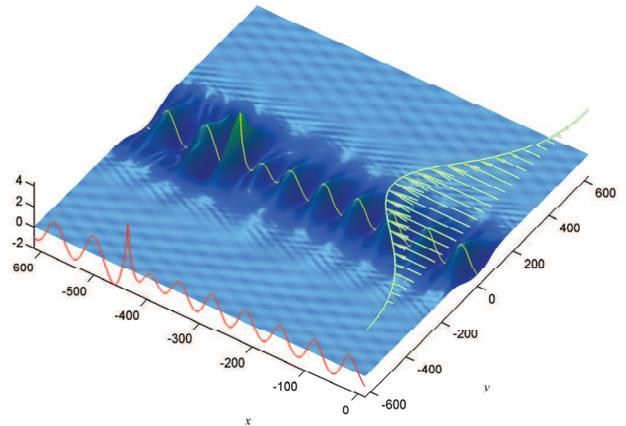}}
\caption{Simulation of propagation of one trapped mode with $k=0.1$~rad/m and $kH/2=0.15$ with a $5$\% initial longitudinal modulation. Surface elevation at time $t=829.4$~s, the maximal wave is characterized by $kH/2 \approx 0.32$. The longitudinal cross-section through the peak of the maximal wave is shown by lines above the current maximum and above the $x$-axis. The profile of the current is shown above the  $y$-axis.}
\label{fig:One_Mode_RogueWave}
\end{figure}

\begin{figure}
\centerline{\includegraphics[width=9.5cm]{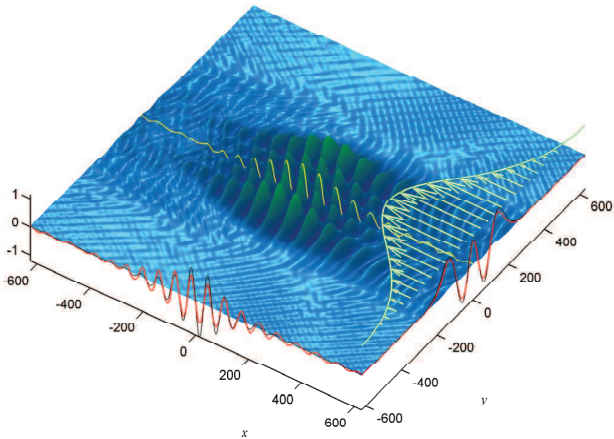}}
\caption{Simulation of a solitary group of trapped waves belonging to the fifth mode (the surface at $t\approx 868$~s is shown, see movie \cite{MovieSoliton5}). The snapshot longitudinal cross-section  is shown by red solid  lines  above the current maximum and above the $x$-axis; the corresponding sections of the initial condition  are shown by thin black curves.  The snapshot  transverse cross-section is shown in front of the surface above the $y$-axis.  }
\label{fig:Soliton_Mode5}
\end{figure}

\emph{Nonlinear dynamics of trapped waves in simulations of the primitive equations.---}
To verify the obtained asymptotic description a few key solutions of (\ref{NLS}) are tested below by means of strongly nonlinear numerical simulations of the primitive hydrodynamic equations.
The High Order Spectral Method (HOSM) to solve the potential Euler equations is adopted for the situations when the four-wave interactions dominate. The approach similar to \cite{Westetal1987} is employed, where the terms responsible for current $U(y)$ have been introduced. 
 The computational domain is periodic in both, $x$ and $y$ coordinates. The current is chosen to be close to $sech^2y$, it is taken to be periodical in $y$ with widely separated humps; it is specified as
$U = U_{max} cn^2{(2K \frac{y}{L_y},s^2)}$,
where $K(s^2)$ is the complete elliptic integral of the first kind, and $U_{max}=-2$~m/s and $s=0.9$ are used. The current varies from zero to $U_{max}$ at $y=0$; it is shown with arrows in Figs.~\ref{fig:One_Mode_RogueWave},~\ref{fig:Soliton_Mode5}.

\emph{1.~Single trapped mode.}
In the first experiment we verify the ability of trapped waves belonging to a single mode to propagate with no noticeable radiation in the fully nonlinear system. The initial condition has the form of a uniform train of 10 Stokes waves with the wavenumber $k=0.1$~rad/m and steepness $kH/2=0.15$ (where $H$ is the trough-to-crest wave height), modulated in the transverse direction according to the fundamental ($n=0$) mode function, $Y_0(y)$. The function $Y_0$ is found by solving   (\ref{2DBVP}) numerically  
(see details in \cite{ShriraSlunyaev2013}). 

The evolution of an initially  uniform wave train of trapped waves is simulated for about 80 wave periods \cite{MovieStokes}. It  propagates steadily with no evidence of significant radiation or structural deformation. The  presence of some small-amplitude ripples is natural since the initial condition is not exactly an one-mode solution. The examination of the instantaneous wave height record gives some clues of two processes which lead to a slow decrease of the trapped wave height: i)~about $10\%$ of wave height is lost during the first  $\approx 10$ wave periods {(we attribute this to the imperfect initial conditions)}; ii)~a longer-term slow trend  resulting in the total loss of about $2\%$  of energy over  the simulated $80$ wave periods is 
 apparently caused by interaction with noise. In other respects the train of trapped waves exhibits  robustness.

The gravity wave angular frequency for the chosen initial condition is $\omega_g = 0.99$~rad/s. The nonlinear correction to the Stokes wave of moderate steepness, $kH/2=0.15$, adds extra  $\approx 1\%$ to the frequency. The solution of the eigenvalue problem (\ref{2DBVP}) yields  
 $\omega_0 = 0.80$~rad/s, while the frequency spectrum of the simulated surface displacement has the maximum at $\omega = 0.80 \pm 0.01$~rad/s. 

\emph{2.~Modulational instability of a single trapped mode leading to a rogue wave pattern.}
A $5\%$ modulation along $Ox$ was applied to the wave train used in the previous simulation to initiate the modulational instability. 
Also, the reference simulation was performed when the current was set  equal to zero, and the train had no modulation in the transverse direction. Supporting the NLSE prediction, the initial modulations grow in both cases with the maximal waves eventually reaching the breaking limit.

The initially modulated train of trapped waves undergoes further localization of wave energy, and the emerging large wave breaks at some instant, which leads to blowing up of the numerical iterations in time. The picture of surface elevation at the moment close to the wave breaking is given in Fig.~\ref{fig:One_Mode_RogueWave}. Due to the factor $\kappa$ in (\ref{NLS}), the evolution governed by the NLSE for trapped waves on currents is slower compared with the 1-d free gravity waves.
Indeed, the curves of instantaneous maximal wave heights versus time in the simulations discussed above may be fitted onto each other, when the time is scaled with factor $0.65$ in the case of trapped waves. This value differs from $\kappa \approx 0.71$ calculated for the chosen profile of the current;  the  discrepancy is most likely due to the inaccuracy in prescribing the single-mode initial conditions for the trapped wave simulation.
%

\emph{3.~Solitary groups of trapped waves.}
Envelope solitons are the most fundamental solutions supported by the 1-d NLSE, they represent asymptotics of the  initial problem with generic localized initial data.  The 2-d generalizations of the NLSE for deep water waves also admit planar envelope soliton solutions, but they are no longer  asymptotics of the  initial problem and  are known to be unstable with respect to long transverse perturbations \cite{ZakharovRubenchik1974} and, hence,  relatively short lived.

To specify the initial condition for the strongly nonlinear simulations we use the exact analytic solution to equation (\ref{NLS}) in the form of an envelope soliton, $A(x,t=0)=A_s sech{(\sqrt{2} k_c^2 A_s x)}$, with the transverse shape prescribed by the modal function, $Y_n(y)$; two cases ($n=0$ and $n=4$) were considered. In the runs the carrier wave has the same longitudinal wave number, $k_c = 0.1$~rad/m, though the intensity is  smaller, $k_c A_s \approx 0.12$.

Fig.~\ref{fig:Soliton_Mode5} shows the result of simulation of a solitary wave which corresponds to the fifth mode. The surface snapshot corresponds to the moment when the solitary group has passed the computational domain twice, that is about $40$ wave lengths. The longitudinal section of the solitary group is shown by lines above the maximum of the current  and above the $x$-axis (red solid line); the transversal section of the group is shown in front of the surface above the $y$-axis (red solid line). These sections are compared with the corresponding sections of the initial condition (thin black curves). The amplitude of the solitary wave group ends up somewhat reduced compared to the initial condition; the radiated wave patterns are discernable in the figure.
The survived intense solitary wave group in Fig.~\ref{fig:Soliton_Mode5} is an indication of  robustness of such groups of trapped waves; although the solitary group produces some radiation each time it interacts with other wave patterns which exist in the simulation domain \cite{MovieSoliton5}. In the course of evolution the group is slowly loosing energy; the total drop of the maximum wave height over  $120$ wave periods is about  $20-25\%$.
The solitary group of waves belonging to the fundamental mode  preserves  energy  better, though some radiation is observed as well \cite{MovieSoliton0}. The decrease of maximum wave height over the same time of simulation is  about
$
10-15\%$. The  patterns causing  radiation by the groups  can be viewed as an artifact of the imperfect choice of the initial conditions, or an unavoidable element  of the imperfect reality the solitary waves are likely to encounter in  nature.

\emph{{Concluding remarks---}}
In the context of oceanic waves the  robust solitary groups of trapped waves found on jet currents represent a unique case of intense patterns of  deep water gravity waves localized in both dimensions, which can propagate over long distances preserving their structure and energy.
We stress that one-dimensionalization of wave dynamics occurs without assumption of a narrow angular spectrum; in the example  shown in Fig.~\ref{fig:Soliton_Mode5} the  width of the spectrum  is of order of the longitudinal wavenumber, $\Delta k_y / k_c \sim 1$.
It makes relevant for oceanic conditions a huge corpus of theoretical and laboratory studies concerned with one-dimensional wave dynamics. As long as the trapped waves are described by the integrable 1-d NLSE to the leading order, all powerfull mathematical techniques and analytic solutions obtained since 1970-es may be applied (see e.g. \cite{Osborne2010} and references therein).
Correspondingly, the same well studied dynamics resulting in extreme waves for the planar waves (including rogue waves in optical fibers, where the NLSE and its solutions may work perfectly well \cite{YeomEggleton2007, Onoratoetal2013}) is also expected for the trapped waves. As a particular conclusion, a higher likelihood of rogue wave events in the field of trapped waves should be expected. Exploiting integrability of the NLSE, elements of deterministic forecasting of oceanic rogue waves may be suggested.

The situations when several/many trapped modes and/or passing modes are present and interact the trapped waves require a dedicated study.
Here we note that the phenomenon of one-dimensionalization of wave dynamics and its implications we discussed are not confined to water waves on currents; similar effects are likely in all branches of physics wherever there are guided waves. 

The work was supported by the EC 7th Framework Grant PIIF-GA-2009-909389. AS acknowledges the support from RFBR via grants 11-02-00483 and 12-05-33087.

\end{document}